\documentclass[12pt]{article}
\usepackage{graphicx}
\usepackage{epsfig}
\usepackage{cite}

\newcommand\pubnumber{~}
\newcommand\pubdate{~}

\def\napoli{Department of Physics\\
University of Cincinnati, Cincinnati OH, USA}
\def\support{\footnote{~}}

\def\Title#1{\begin{center} {\Large #1 } \end{center}}
\def\Author#1{\begin{center}{ \sc #1} \end{center}}
\def\Address#1{\begin{center}{ \it #1} \end{center}}

\newcommand\pubblock{\rightline{\begin{tabular}{l} \pubnumber\\
         \pubdate  \end{tabular}}}
\newenvironment{Abstract}{\begin{quotation}  }{\end{quotation}}
\newenvironment{Presented}{\begin{quotation} \begin{center} 
             PRESENTED AT\end{center}\bigskip 
      \begin{center}\begin{large}}{\end{large}\end{center} \end{quotation}}
\def\Acknowledgements{\bigskip  \bigskip \begin{center} \begin{large}
             \bf ACKNOWLEDGEMENTS \end{large}\end{center}}

\def\beq{\begin{equation}}
\def\eeq#1{\label{#1}\end{equation}}
\def\eeqn{\end{equation}}

\def\beqa{\begin{eqnarray}}
\def\eeqa#1{\label{#1}\end{eqnarray}}
\def\eeqan{\end{eqnarray}}

\let\bar=\overbar

\def\Dslash{\not{\hbox{\kern-4pt $D$}}}
\def\dslash{\not{\hbox{\kern-2pt $\del$}}}

\def\msb{{\bar{\ssstyle M \kern -1pt S}}}

\begin{document}
\begin{titlepage}
\pubblock

\vfill
\Title{Summary of $CP$ Violation in $D\rightarrow hh$ Decays at Belle}
\vfill
\Author{ Eric White, for the Belle Collaboration \support}
\Address{\napoli}
\vfill
\begin{Abstract}
Although $CP$ violation is predicted to be small in the Standard Model, there exists a continuing theoretical interest within the charm sector due to enhancements of CPV through new physics processes.
We present an overview of the measurements of direct $CP$ violation in the charm sector at Belle. 
We give a brief review of the theory of $CP$ violation in charm physics, then discuss recent measurements for the decay modes
$D^{+}_{(s)}\rightarrow \phi\pi^{+}$;
$D^{0}\rightarrow K^{0}_{S}\pi^{0}, K^{0}_{S}\eta^{(\prime)}$;
$D^{+}\rightarrow \pi^{+}\eta^{(\prime)}$;
and $D^{+}\rightarrow  K^{0}_{S}\pi^{+}$.
\end{Abstract}
\vfill
\begin{Presented}
The $5^{\textrm{th}}$ International Workshop on Charm Physics\\
Charm 2012 \\
Honolulu, Hawai'i 96822, USA,  May 14--17, 2012
\end{Presented}
\vfill
\end{titlepage}
\def\thefootnote{\fnsymbol{footnote}}
\setcounter{footnote}{0}

\section{Introduction}

The violation of charge-parity ($CP$) symmetry has been well-established in the neutral $K^{0}$ and $B^{0}$ systems~\cite{Christ,BaBar,Belle,PDG}, while a first measurement of $CP$ violation (CPV) in the $B^{0}_{s}$ system has been reported by the LHCb collaboration~\cite{LHCb}.
Along with the observation of $D^{0}$ mixing~\cite{BaBarMix,BelleMix}, 
there has been a renewal of  interest in charm physics beyond the SM.
CPV is described in the Standard Model (SM) by the Kobayashi-Maskawa phase in the Cabibbo-Kobayashi-Maskawa (CKM) mixing matrix~\cite{CKM1,CKM2}. 
However, since the expected size of CPV is not sufficient to account for the known matter-antimatter imbalance in our Universe,
it is believed that additional sources of $CP$ asymmetry could be generated by physics beyond the Standard Model~\cite{Uni}.

Although quark mixing has been measured to a level that is consistent with theory, recent evidence for CPV
in singly Cabibbo-suppressed $D^{0}$ decays~\cite{LHCb} has renewed theoretical interest in charm
physics.
 Prior to the measurement by LHCb, CP asymmetries in $D^{0}$ decays to two-body final states were expected to be very small in the SM~\cite{SM1,SM2,SM3,SM4}, of the order $O(10^{-3})$.
Thus, it is necessary to perform experimental searches in other charm decays such as $D^{0}\rightarrow K^{+}K^{-}$,
 as well as other two-body final states $hh$.

The precise theoretical determination of CPV in the charm sector is difficult to achieve,
since the charm quark is too heavy for chiral perturbation theory and yet too light for the reliable application of heavy-quark effective theory.
Thus, the determination of whether or not an observed effect is a clear sign of new physics beyond
the SM is rather challenging~\cite{GIM1,GIM2,GIM3,GIM4}. 
In order to clarify this phenomenological picture, it is crucial to study CPV in other charm decays.
In this paper. we present recent measurements by the Belle collaboration for the decay modes
$D^{+}_{(s)}\rightarrow \phi\pi^{+}$;
$D^{0}\rightarrow K^{0}_{S}\pi^{0}, K^{0}_{S}\eta^{(\prime)}$;
$D^{+}\rightarrow \pi^{+}\eta^{(\prime)}$;
and $D^{+}\rightarrow  K^{0}_{S}\pi^{+}$.

\section{The Belle Detector}

The data sample was collected by the Belle detector~\cite{belle_detector} located at the KEKB asymmetric $e^+e^-$
collider operating at or near the $\Upsilon(4S)$ resonance.
The Belle detector is a large-solid-angle magnetic spectrometer consisting of a silicon vertex detector (SVD), 
a 50-layer central drift chamber (CDC), an array of aerogel threshold 
\v{C}erenkov counters (ACC), a barrel-like arrangement of time-of-flight scintillation counters (TOF), 
and an electromagnetic calorimeter (ECL) based on CsI(Tl) crystals.
These detector elements are all located inside a superconducting solenoid 
coil that provides a 1.5 T magnetic field. For charged hadron 
identification, a likelihood ratio is formed based on $dE/dx$ 
measured in the CDC and the response of the ACC and TOF.   
Charged kaons are identified using a likelihood requirement that is approximately 86\% efficient for $K^\pm$ and has a $\pi^\pm$ 
misidentification rate of about 8\%.
A more detailed description of the detector can be found elsewhere~\cite{belle2}.


\section{Formalism}

The time-integrated $CP$ asymmetry for a $D^{0}$ meson decaying to some final state $f$ is defined as
\begin{equation}
A_{CP}^{D\rightarrow f} \equiv \frac{\Gamma(D \rightarrow f) - \Gamma(\overline{D} \rightarrow \overline{f})}{\Gamma(D \rightarrow f) + \Gamma(\overline{D} \rightarrow \overline{f})},
\end{equation}
where $\Gamma$ is the partial decay width for the process and $\overline{D}$, $\overline{f}$ are the charge-conjugate states.
However, for any given mode the reconstructed asymmetry $A_{rec}$ can include contributions from sources other than $A_{CP}$, such as kinematics and detector asymmetries.

For an asymmetric detector such as Belle, the production of $D$ mesons through the decay $e^{+}e^{-}\rightarrow c\bar{c}$ will contain a forward-backward asymmetry $A_{FB}$ due to interferences between electroweak ($Z^{0}$) and electromagnetic ($\gamma$)  processes, as well as higher-order QED effects.
In final states containing a neutral kaon, SM mixing between $K^{0}-\bar{K}^{0}$ can contribute a small $CP$ asymmetry even in the absence of $CP$ violation in the charm decay.
Additional contributions to the reconstructed asymmetry can also arise from detection asymmetries.

\section{$D^{+}_{(s)}\rightarrow \phi\pi^{+}$}

The largest contributions to $CP$ violation in the Standard Model occur through singly Cabibbo-suppressed (SCS) decays, such as $D^{+}\rightarrow \phi\pi^{+}$~\cite{GIM1}.
In order to cancel detection asymmetries and other systematic/experimental effects,  the measured quantity is constructed as the difference in reconstruction asymmetries between the $D^{\pm}$ and $D^{pm}_{s}$:
\begin{equation}
\Delta A_{rec} = \frac{N(D^{+}) - N(D^{-})}{N(D^{+}) + N(D^{-})} - \frac{N(D_{s}^{+}) - N(D_{s}^{-})}{N(D_{s}^{+}) + N(D_{s}^{-})},
\end{equation}
where the second term corresponds to the Cabibbo-favored (CF) decay $D_{s}^{+}\rightarrow \phi\pi^{+}$.
Since this latter decay proceeds through the CKM matrix elements $V_{cs}V^{\star}_{ud}$ \textemdash and thus is expected to have negligible contributions to $A_{CP}$ \textemdash the quantity $\Delta A_{rec}$ directly probes $A_{CP}$ in $D^{+}\rightarrow \phi\pi^{+}$ decays~\cite{Neg}.

The measurement is based on 955 fb$^{-1}$ of data collected at the $\Upsilon(4S)$ resonance, with a fraction of the data recorded at the $\Upsilon(1S)$, $\Upsilon(2S)$, $\Upsilon(3S)$, and $\Upsilon(5S)$ resonances.
The $\phi$ meson is reconstructed through the $\phi\rightarrow K^{+}K^{-}$ decay mode. 
The reconstruction asymmetry for this decay can be expressed as the sum of several small contributions:
\begin{equation}\label{eq:phirec}
A_{rec} = A_{CP} + A_{FB}(\cos\theta^{star}) + A_{\epsilon}^{KK} + A_{\epsilon}^{\pi}(p_{\pi},\cos\theta_{\pi}).
\end{equation}
The forward-backward asymmetry is an odd function of the cosine of the $D$ meson's polar angle in the CM frame and could in principle differ between the $D^{+}$ and $D^{+}_{s}$ due to fragmentation effects.
There can be additional contributions to the asymmetry due to differences in the reconstruction efficiencies of oppositely charged kaons ($A_{\epsilon}^{KK}$) and pions ($A_{\epsilon}^{\pi}$).

The total number of positively and negatively charged $D$ mesons can be calculated by integrating over the intrinsic laboratory phase-space distribution of the kaon pair convolved with the detection efficiency.
Assuming the efficiencies are proportional to $(1 \pm A_{\epsilon}^{K}(x))$ and neglecting higher-order terms in $A_{\epsilon}^{K}$, the kaon efficiency asymmetry can be written as
\begin{equation}\label{eq:akint}
A_{\epsilon}^{KK} = \int (P_{1}(x) - P_{2}(x))A_{\epsilon}^{K}(x) dx,
\end{equation}
where $P_{1}(x)$ and $P_{2}(x)$ are the normalized distributions of detected same-sign and opposite-sign kaons, respectively.
Here, the integration runs over the kaon phase space $x \equiv (p, \cos\theta)$.
The effiency asymmetry $A_{\epsilon}^{\pi}$ depends on the pion momentum and polar angle in the laboratory frame.
For the difference in reconstruction asymmetries $A_{rec}$ measured in bins of three-dimensional (3D) phase space $(\cos\theta^{\star}, p_{\pi}, \cos\theta_{\pi})$, the pion efficiency asymmetry $A_{\epsilon}^{\pi}$ cancels.
The 3D phase space is divided into $10 \times 10 \times 10$ bins of equal size with $p_{\pi} < 5$ GeV/$c$.
In each bin, the $D^{\pm}$ and $D^{\pm}_{s}$ yields and asymmetry difference $\Delta A_{rec}$ are measured using a binned likelihood fit to the $M_{KK\pi}$ distributions.
The sum of the mass distributions correspond contain 237525 $\pm$ 577 $D^{\pm}$ and 722871 $\pm$ 931 $D^{\pm}_{s}$ decays.
The asymmetry difference in each bin is then corrected with $\Delta A_{\epsilon}^{KK}$, so that
\begin{equation}
\Delta A_{rec}^{cor} = \Delta A_{rec} - \Delta A_{\epsilon}^{KK}.
\end{equation}
The corrections are then obtained by using Eq.~\ref{eq:akint} and the distributions for $P_{1}(x)$ and $P_{2}(x)$ in data.

The kaon asymmetry $A_{\epsilon}^{K}$, used in Eq.~\ref{eq:akint}, is measured with the help of $D^{0}\rightarrow K^{-}\pi^{+}$ decays.
The measured asymmetry for these decays can be written as $A_{rec} = A_{CP} + A_{FB} - A_{\epsilon}^{K} + A_{\epsilon}^{\pi}$.
By assuming negligible $CP$ violation in CF decays and the same forward-backward asymmetry,
and by neglecting the $A_{\epsilon}^{KK}$ in Eq.~\ref{eq:phirec},
the measured asymmetry difference is equal to $A_{\epsilon}^{K}$~\cite{Diff}.

The asymmetry of $D^{+}_{s}\rightarrow \phi\pi^{+}$ is measured using fitted yields in the aforementioned 3D bins.
This asymmetry map is then used in conjuction with a sideband subtraction technique to weight $D^{0} \rightarrow K^{-}\pi^{+}$ events and determine the corrected yields for $D^{0}$ and $\overline{D}^{0}$ in bins of kaon phase space.
The following corrections are obtained for the total 3D phase space:
$A_{\epsilon}^{KK} = (0.060 \pm 0.013)\%$ for $D^{+}$,
 $A_{\epsilon}^{KK} = (-0.051 \pm 0.012)\%$ for $D^{+}_{s}$,
and $\Delta A_{\epsilon}^{KK} = (0.111 \pm 0.025)\%$.
The non-zero difference is due to momentum asymmetries in $D^{+}$ and $D^{+}_{s}$ decays.

The corrected asymmetry differences in 3D bins are used calculate the error-weighted averages in bins of $\cos\theta^{\star}$ using a least-squares fit.
The quantities $A_{CP}$ and $\Delta A_{FB}$  are then extracted by adding/subtracting the asymmetry difference in opposite bins of $\cos\theta^{\star}$ using the following relations:
\begin{eqnarray}
A_{CP} &=& \frac{\Delta A_{rec}^{cor}(\cos\theta^{\star}) + \Delta A_{rec}^{cor}(-\cos\theta^{\star})}{2}, \\
\Delta A_{FB} &=& \frac{\Delta A_{rec}^{cor}(\cos\theta^{\star}) - \Delta A_{rec}^{cor}(-\cos\theta^{\star})}{2}
\end{eqnarray}

\begin{figure}
\begin{center}
\begin{tabular}{cc}
\epsfig{file=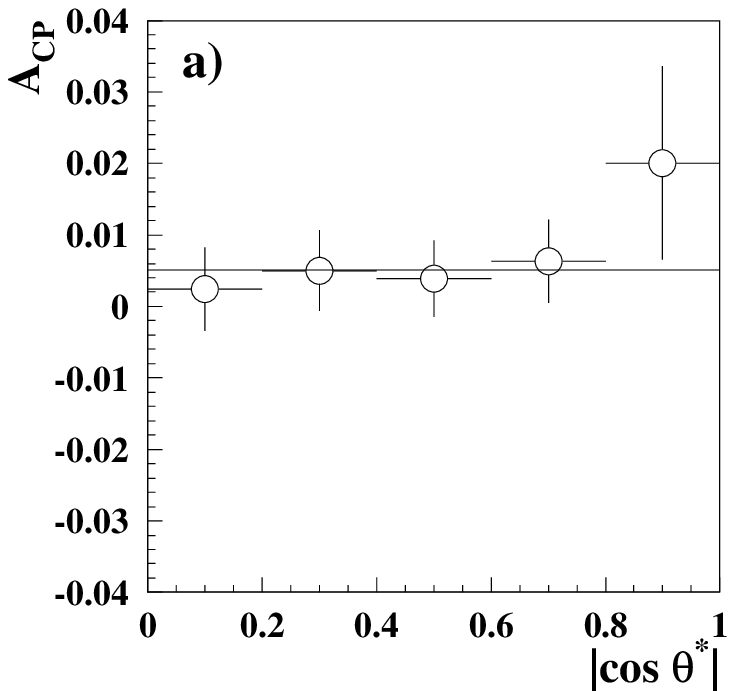,width=0.45\linewidth} &
\epsfig{file=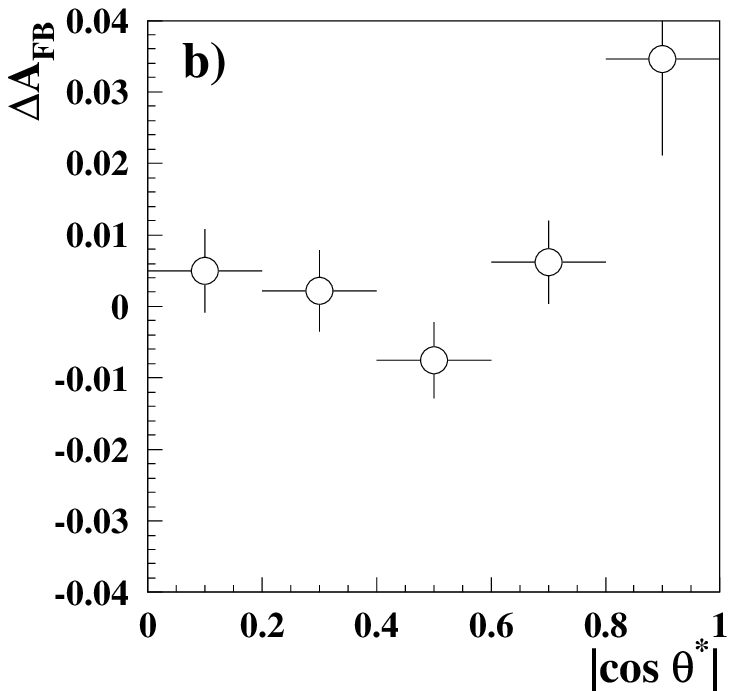,width=0.45\linewidth}
\end{tabular}
\caption{$CP$-violating asymmetry $A_{CP}$ (left) and forward-backward asymmetry difference $\Delta A_{FB}$ (right), in bins of $|\cos\theta^{\star}|$. The horizontal line in the plot for $A_{CP}$ represents the value of a constant fit to the data points.}
\label{fig:phiCP}
\end{center}
\end{figure}

 The fitted $CP$ asymmetry is found to be $A_{CP} = (0.51 \pm 0.28 \pm 0.05)\%$,
where the first error is statistical and the second systematic~\cite{anal1}.
The result is consistent with no $CP$ violation within 1.8 standard deviations and agrees with SM predictions,
and corresponds to an improvement of five times the precision of previous measurements by CLEO~\cite{CLEO} and BaBar~\cite{BabX}.
Fig.~\ref{fig:phiCP} shows the measured difference in forward-backward asymmetries, with the $\chi^{2}$ test of fitted values corresponding to a confidence level of 6\%. 
This corresponds to a first measurement of the difference in forward-backward asymmetries between the $D^{+}$ and $D^{+}_{s}$ mesons; no significant deviation from zero was found. 


\section{$D^{0}\rightarrow K^{0}_{S}\pi^{0}$ and $D^{0}\rightarrow K^{0}_{S}\eta^{(\prime)}$}

Observed $K^{0}_{S}P^{0}$ final states are mixtures of CF $D^{0}\rightarrow \overline{K}^{0}P^{0}$ and DCS $D^{0}\rightarrow K^{0}P^{0}$ decays.
Standard Model $CP$ violation in these processes is generated from mixing and interference of decays with and without mixing,
often parameterized by $a^{ind}$.
The asymmetry from $K^{0}-\overline{K}^{0}$ mixing is measured to be $A^{K^{0}}_{CP} = (-0.332 \pm 0.006)\%$~\cite{PDG} and arises in the value of $A_{CP}$ for $D^{0}\rightarrow K^{0}_{S}P^{0}$, when contributions from DCS decays are ignored.
It is estimated that for new physics processes containing additional weak phases other that of the Kobayashi-Maskawa phase,
interferences between CF and DCS channels could generate direct $CP$ asymmetries on the order of 1\% for $D^{0}\rightarrow K^{0}_{S}P^{0}$ decays~\cite{inter}.
Thus, an observation of $A_{CP}$ that is inconsistent with $A^{K^{0}}_{CP}$ would provide strong evidence for physics processes beyond the SM~\cite{NP}.

The flavor of the neutral $D$ meson is identified using the charge of the low-momentum (``soft'') pion $\pi^{+}_{s}$ in the decay $D^{\star +}\rightarrow D^{0}\pi^{+}_{s}$.
The $CP$ asymmetry is then measured from the reconstruction asymmetry, defined as
\begin{equation}
A_{rec} = \frac{ N_{rec}^{D^{0}\pi^{+}_{s}} - N_{rec}^{\overline{D}^{0}\pi^{-}_{s}} }{ N_{rec}^{D^{0}\pi^{+}_{s}} + N_{rec}^{\overline{D}^{0}\pi^{-}_{s}} },
\end{equation}
where $N_{rec}$ is the number of reconstructed decays for each flavor.
Other than a forward-backward asymmetry $A_{FB}$, the measured asymmetry includes an additional contribution from the detection efficiency asymmetry $A_{\epsilon}^{\pi_{s}}$ between positively and negatively charged soft pions.
Since neutral kaons are reconstructed with $\pi^{+}\pi^{-}$ combinations and $P^{0}$ with either $\gamma\gamma$ or $\gamma\gamma\pi^{+}\pi^{-}$ final states, asymmetries in the detection of $K^{0}_{S}$ and $P^{0}$ cancel.
The reconstruction asymmetry can than be written as
\begin{equation}
A_{rec} = A_{CP} + A_{FB}(\cos\theta_{D^{\star}}) + A_{\epsilon}^{\pi_{s}}(p^{T}_{\pi_{s}},\cos\theta_{\pi_{s}}),
\end{equation}
where $A_{FB}$ is an odd function of the cosine of the polar angle of $D^{\star +}$ in the CMS frame,
$A_{\epsilon}^{\pi_{s}}$ depends on the transverse momentum and polar angle of the slow pion in the laboratory frame,
and $A_{CP}$ is independent of all kinematic variables.

By assuming the same $A_{FB}$ for $D^{\star+}$ and $D^{0}$ mesons,
the decays $D^{\star+}\rightarrow D^{0}\pi^{+}_{s}\rightarrow  K^{-}\pi^{+}\pi^{+}_{s}$ (tagged) and  $D^{0}\rightarrow K^{-}\pi^{+}$ (untagged) can be used to correct for $A_{\epsilon}^{\pi_{s}}$.
This is accomplished by subtracting the measured asymmetries in these two decay modes, $A_{rec}^{untagged}$ and $A_{rec}^{tagged}$.
Denoting the reconstruction efficiency corrected for $A_{\epsilon}^{\pi_{s}}$ as $A^{D^{0}\pi^{+}_{s}}_{rec,corr}$,
the quantities $A_{CP}$ and $\Delta A_{FB}$ can then extracted by adding/subtracting the corrected asymmetries using the following relations:
\begin{eqnarray}
A_{CP} &=& \frac{A^{D^{0}\pi^{+}_{s}}_{rec,corr}(\cos\theta_{D^{\star+}}) + A^{D^{0}\pi^{+}_{s}}_{rec,corr}(-\cos\theta_{D^{\star+}})}{2} \\
A_{FB} &=& \frac{A^{D^{0}\pi^{+}_{s}}_{rec,corr}(\cos\theta_{D^{\star+}}) - A^{D^{0}\pi^{+}_{s}}_{rec,corr}(-\cos\theta_{D^{\star+}})}{2}.
\end{eqnarray}

The measurement is based on 791 fb$^{-1}$ of data collected at or near the $\Upsilon(4S)$ resonance.
The mass difference $M(D^{\star}) - M(D)$ is reconstructed for each mode and parameterized as a sum of Gaussian and bifurcated Gaussians with common mean.
The asymmetry of the negatively and positively charged $D^{\star}$ yields, as well as their sum, is obtained directly from simultaneous fit to the distribution of each candidate.
Table~\ref{tab:modes} lists the results of these fits.
\begin{table}[t]
\begin{center}
\begin{tabular}{lcc}  
~ &  $N_{S}$ &  $A_{rec}$ (\%) \\ \hline
$D^{\star+}\rightarrow D^{0}\pi^{+}_{s}\rightarrow K^{0}_{S}\pi^{0}\pi^{+}_{s}$  &  $326303 \pm 679$   &   $+0.19 \pm 0.19$  \\
$D^{\star+}\rightarrow D^{0}\pi^{+}_{s}\rightarrow K^{0}_{S}\eta\pi^{+}_{s}$  &  $45831 \pm 283$   &   $+1.00 \pm 0.51$  \\
$D^{\star+}\rightarrow D^{0}\pi^{+}_{s}\rightarrow K^{0}_{S}\eta^{\prime}\pi^{+}_{s}$  &  $26899 \pm 211$   &   $+1.47 \pm 0.67$  
\end{tabular}
\caption{The sum $N_{S}$ and asymmetry $A_{rec}$ for $D^{\star+}$ and $D^{\star-}$ yields from the fits. Uncertainties are statistical only. Systematic uncertainties for final results are listed at the end of this section.}
\label{tab:modes}
\end{center}
\end{table}

\begin{figure}
\begin{center}
\begin{tabular}{c}
\epsfig{file=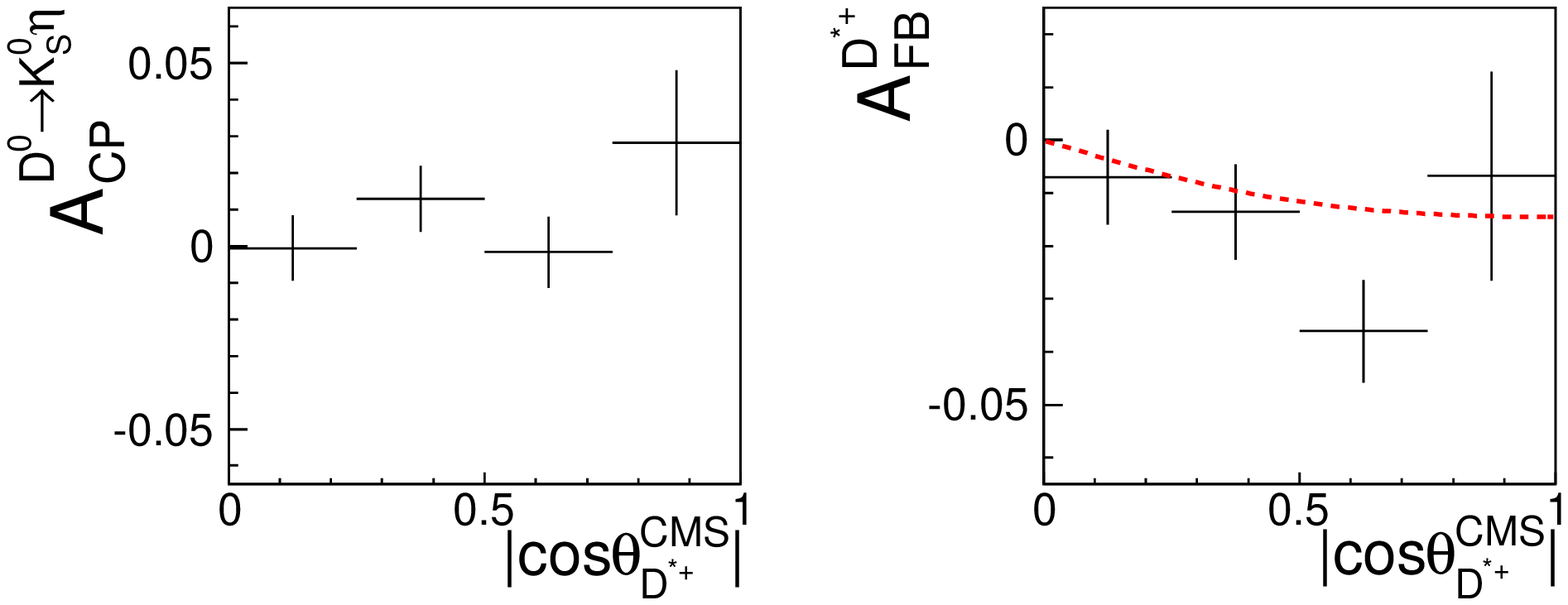,width=0.7\linewidth} \\
\epsfig{file=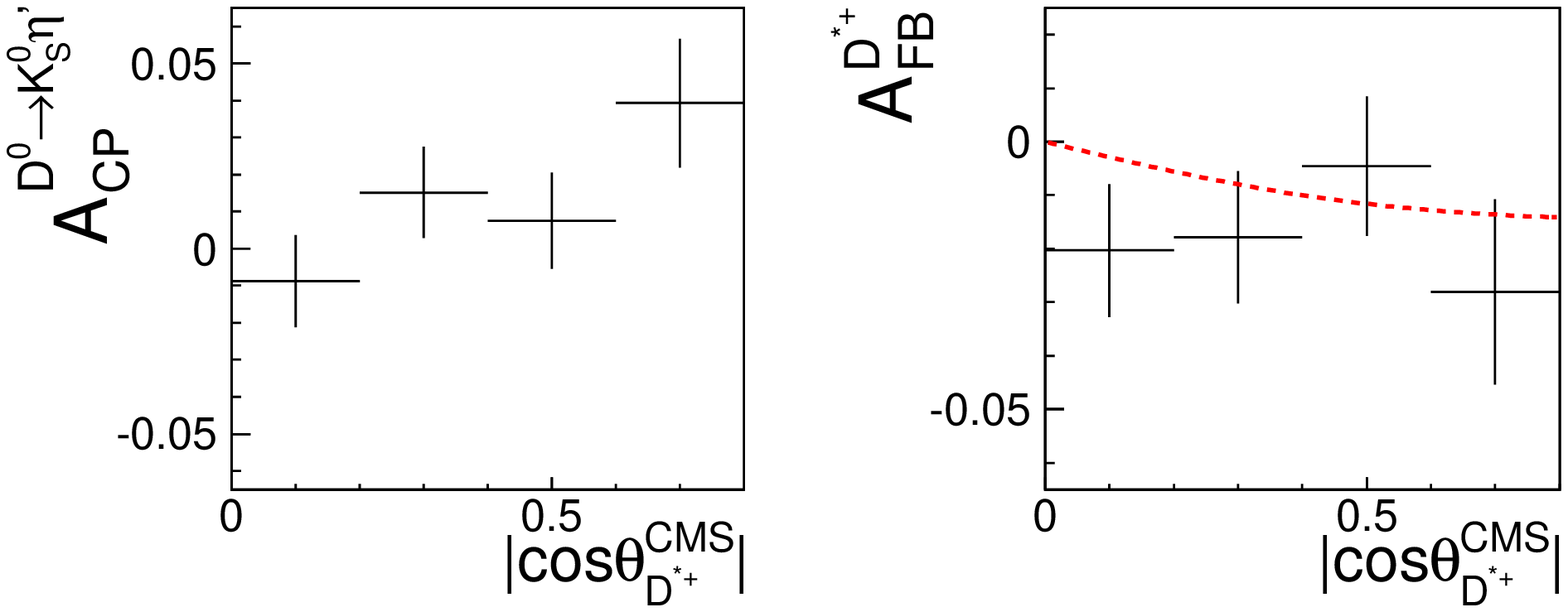,width=0.7\linewidth} \\
\epsfig{file=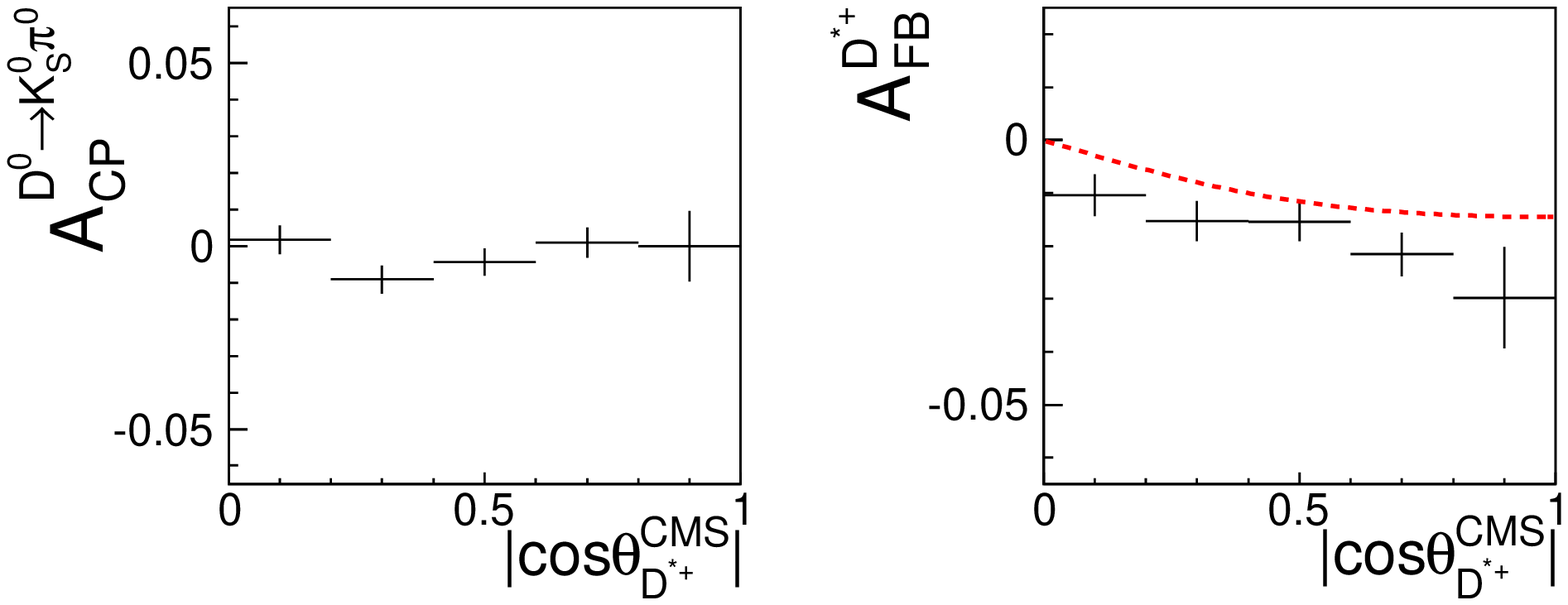,width=0.7\linewidth} 
\end{tabular}
\caption{Measured $A_{CP}$ (left) and $A_{FB}$ (right) values as a function of $|\cos\theta_{D^{\star+}}|$.
Top plots are for $K^{0}_{S}\pi^{0}$, middle plots for $K^{0}_{S}\eta$, and bottom plots for $K^{0}_{S}\eta^{\prime}$ final states. The dashed curves show the leading-order prediction for $A_{FB}$.}
\label{fig:etaCP}
\end{center}
\end{figure}

The values of $A_{\epsilon}^{\pi_{s}}$ are extracted from simultaneous fits to the $M(D)$ distributions, with similar parameterization.
An untagged sample of $D^{0}\rightarrow K^{-}\pi^{+}$ is used to evaluate $A_{rec}$ in bins of transverse momentum ($p_{TD^{0}}$) and polar angle ($\cos\theta_{D^{0}}$).
Factors of $(1 \mp A_{rec})$ are used to weight tagged $D^{\star+}\rightarrow D^{0}\pi^{+}_{s}\rightarrow  K^{-}\pi^{+}\pi^{+}_{s}$ events for each $D^{\star\pm}$ candidate,
and the values of $A_{\epsilon}^{\pi_{s}}$ are then extracted.
After dividing the data samples into bins of momentum and polar angle for the slow pion in the CM frame and correcting each $D^{\star}$ event by the corresponding factor of $A_{\epsilon}^{\pi_{s}}$,
the values of $A_{FB}$ and $A_{CP}$ are then fitted.
Fig.~\ref{fig:etaCP} shows $A_{CP}$ and $A_{FB}$ as a function of $|\cos\theta_{D^{\star+}}|$.

The values obtained for the $CP$ asymmetry are $A_{CP}^{K_{S}^{0}\pi^{0}} = (-0.28 \pm 0.19 \pm 0.10)\%$,
$A_{CP}^{K_{S}^{0}\eta} = (+0.54 \pm 0.51 \pm 0.16)\%$, and $A_{CP}^{K_{S}^{0}\eta^{\prime}} = (-0.98 \pm 0.67 \pm 0.14)\%$~\cite{anal2}, where the first uncertainties are statistical and the second are systematic.
No evidence for $CP$ violation is observed. The results are consistent with SM predictions and currently serve as the most stringent constraint for new physics models arising from $CP$ violation in the charm sector.
The measurement of the decay $D^{0} \rightarrow K^{0}_{S}\pi^{0}$ is one of the most precise measurements of any $CP$ asymmetry in the charm sector.
For $D^{0} \rightarrow K^{0}_{S}\eta$ and  $D^{0} \rightarrow K^{0}_{S}\eta^{\prime}$, the results presented represent the first measurement of $CP$ violation in these modes.


\section{$D^{+}\rightarrow K^{+}\eta^{(\prime)}$ and $D^{+}\rightarrow \pi^{+}\eta^{(\prime)}$ }

Decays of charged $D$ mesons are important to understanding sources of SU(3) flavor symmetry breaking, as well as the nature of $CP$ violation that arises in the CKM flavor-mixing matrix in the Standard Model~\cite{charm1,charm2}.
The two-body decays containing an $\eta^{(\prime)}$ are either doubly Cabibbo-suppressed ($K^{+}\eta^{(\prime)}$) or singly Cabibbo-suppressed ($\pi^{+}\eta^{(\prime)}$).
With current experimental sensitivity, observation of $CP$ violation in two-body decays of a charged $D$ meson containing an $\eta^{(\prime)}$ would represent strong evidence of new physics processes beyond the Standard Model.

\begin{figure}
\begin{center}
\begin{tabular}{cc}
\epsfig{file=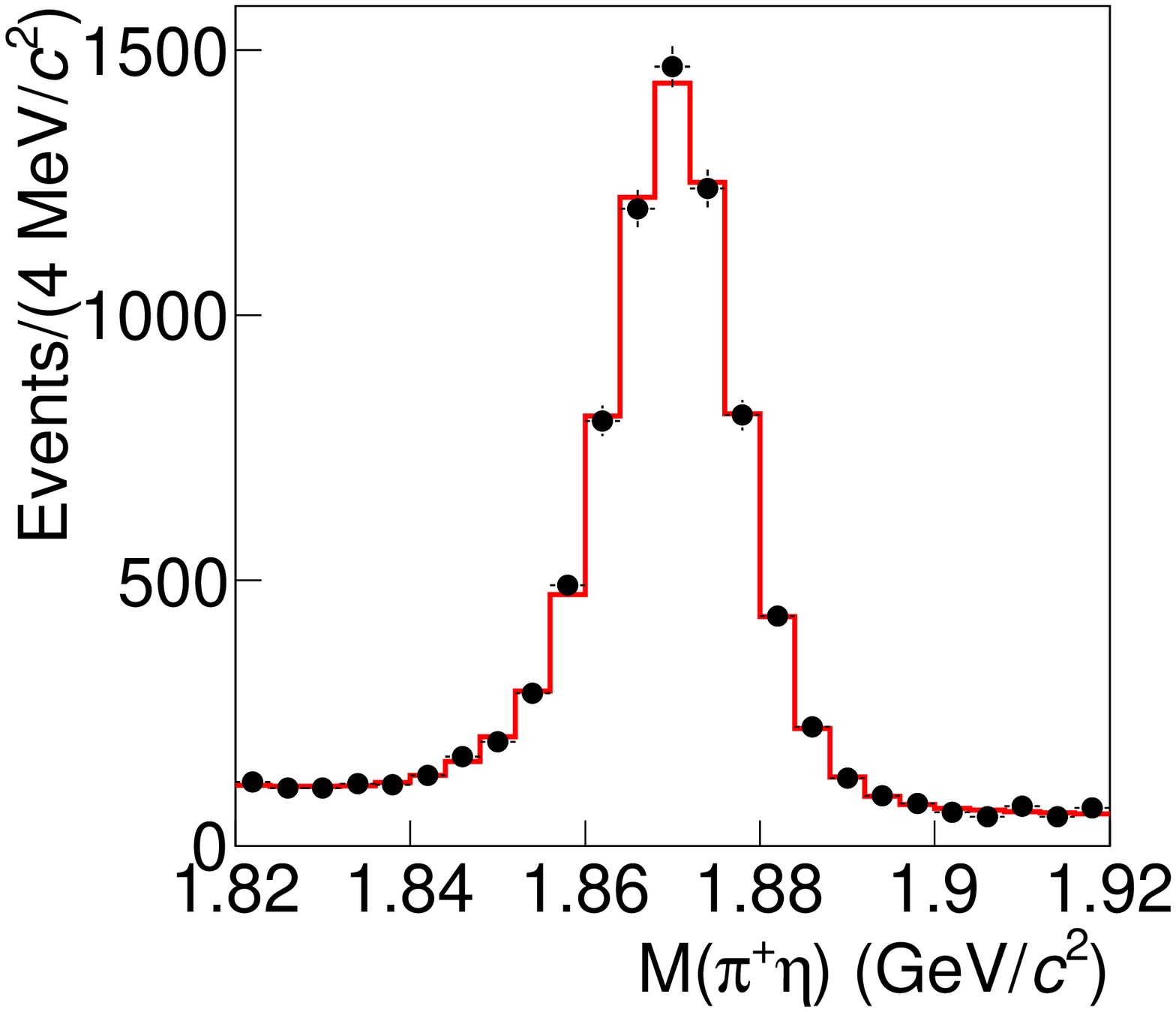,width=0.45\linewidth} &
\epsfig{file=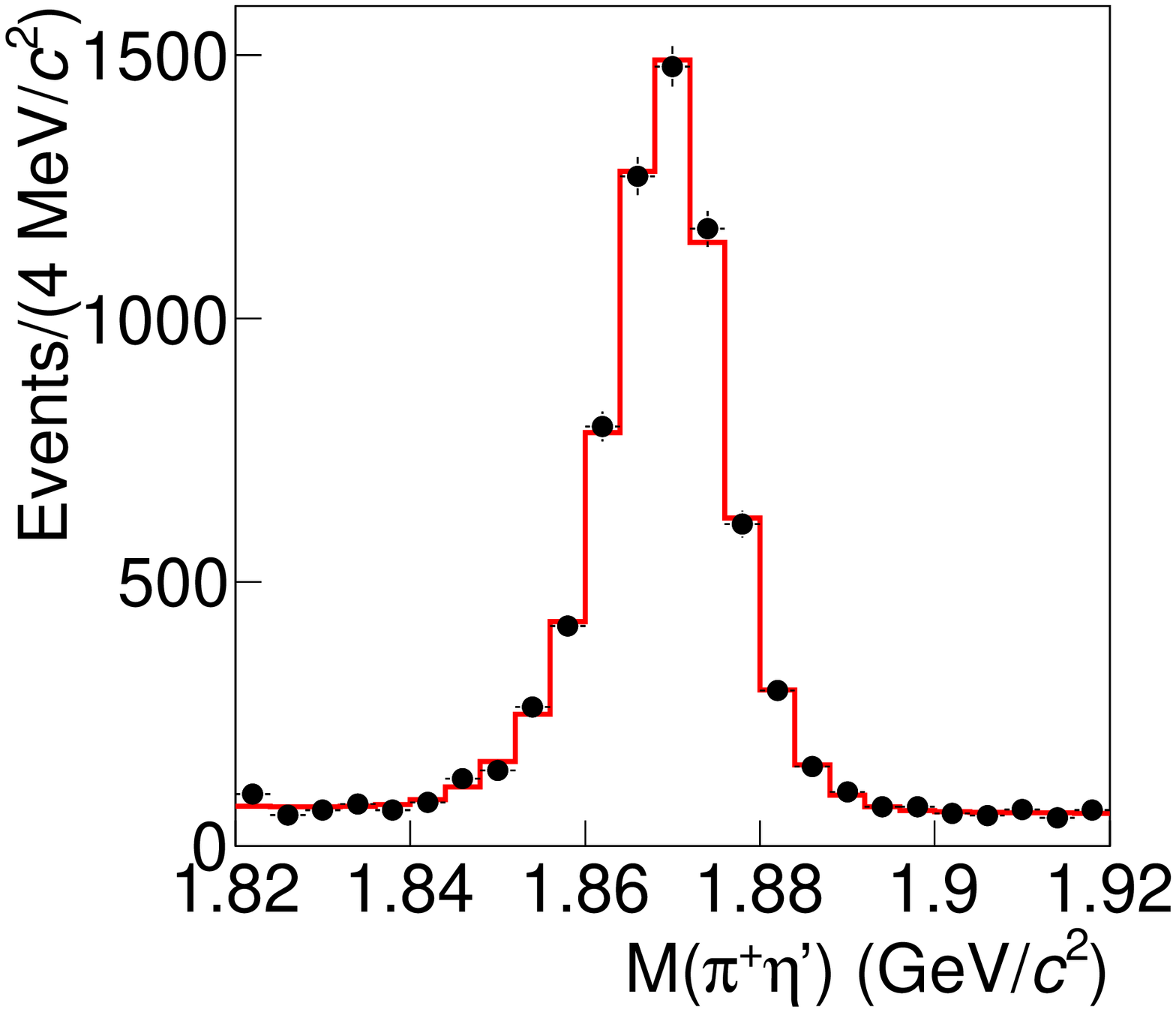,width=0.45\linewidth} 
\end{tabular}
\caption{The invariant mass distributions used for the branch-
ing fraction measurements, corresponding to $\pi^{+}\eta$ (left) and $\pi^{+}\eta^{\prime}$ (right) final states. Points with error bars and histograms correspond to the data and the fit, respectively.}
\label{fig:etapiCP}
\end{center}
\end{figure}

 The data used for this measurement corresponds to an integrated luminosity of 791 fb$^{-1}$, collected at or near the $\Upsilon(4S)$ resonance.
The $\eta$ is reconstructed through the $\eta^{(\prime)}\rightarrow\pi^{+}\pi^{-}\pi^{0}$ decay mode, 
while $\eta^{\prime}$ is reconstructed through the $\pi^{+}\pi^{-}\eta_{\gamma\gamma}$ decay.
Fig.~\ref{fig:etapiCP} shows the invariant mass distributions for $\pi^{+}\eta$ and $\pi^{+}\eta^{\prime}$ after event selection.

The reconstruction asymmetry for this decay can be expressed as:
\begin{equation}\label{eq:phirec}
A_{rec} = A_{CP}^{\pi^{+}\eta^{\prime}} + A_{FB}^{D^{+}} + A_{\epsilon}^{\pi},
\end{equation}
where the total reconstruction asymmetry includes additional contributions from the forward-backward asymmetry of the charged $D$, as well as the detection effiiciency asymmetry between positively and negatively charged pions.
These additional asymmetries are corrected using a sample of Cabibbo-favored $D^{+}_{s}\rightarrow \phi\pi^{+}$ decays,
in which the $CP$ asymmetry predicted by the SM is expected to be negligible.
Assuming that $A_{FB}$ is the same for charged $D$ and $D_{s}$ mesons,
the $CP$ asymmetry can be extracted by taking the difference between the reconstruction asymmetries of $D^{+}\rightarrow \pi^{+}\eta^{(\prime)}$ and $D^{+}_{s}\rightarrow \phi\pi^{+}$ decays.
The measurement is performed in 3D bins of $p_{T\pi}$ and $\cos\theta_{\pi}$ in the laboratory frame and polar angle $\cos\theta_{D^{+}_{(s)}}$ in the CM frame.
After subtracting the reconstruction asymmetry for $D^{+}_{s}\rightarrow \phi\pi^{+}$,
weighted averages of the $CP$ asymmetry are summed over each bin to give $A_{CP} = (+1.74 \pm 1.13 \pm 0.19)\%$ for $D^{+}\rightarrow \pi^{+}\eta$  and $(-0.12 \pm 1.12 \pm 0.17)\%$ for $D^{+}\rightarrow \pi^{+}\eta^{(\prime)}$~\cite{anal3}.

To date, these measurements correspond to the most precise determinations of $A_{CP}$ in $D^{+}\rightarrow \pi^{+}\eta^{(\prime)}$ decays.
Also measured were the ratios of branching fractions of DCS and SCS modes, for which $\mathcal{B}(D^{+}\rightarrow K^{+}\eta)/\mathcal{B}(D^{+}\rightarrow \pi^{+}\eta) = (3.06 \pm 0.43 \pm 0.14)\%$ and $\mathcal{B}(D^{+}\rightarrow K^{+}\eta^{\prime})/\mathcal{B}(D^{+}\rightarrow \pi^{+}\eta^{\prime}) = (3.77 \pm 0.39 \pm 0.10)\%$.
From these branching fraction ratios, along with that of $D^{0}\rightarrow K^{+}\pi^{-}$~\cite{kpiphase} and the tree and annihilation amplitude relations given in Ref.~\cite{tree}, the relative final-state phase difference between tree and annihilation amplitudes in $D^{+}$ decays is measured to be $\delta_{TA} = (72 \pm 9)^{\circ}$ or $(288 \pm 9)^{\circ}$.


\section{$D^{+}\rightarrow  K^{0}_{S}\pi^{+}$}

To date, the the world average of the $CP$ asymmetry in the decay $D^{+}\rightarrow  K^{0}_{S}\pi^{+}$ is $(-0.54 \pm 0.14)\%$~\cite{PDG}, consistent with the expected asymmetry due to the neutral kaon in the final state.
The $CP$ asymmetry in this decay is defined as
\begin{equation}
A_{CP} \equiv \frac{\Gamma(D^{+}\rightarrow  K^{0}_{S}\pi^{+}) - \Gamma(D^{-}\rightarrow  K^{0}_{S}\pi^{-})}{\Gamma(D^{+}\rightarrow  K^{0}_{S}\pi^{+}) + \Gamma(D^{-}\rightarrow  K^{0}_{S}\pi^{-})}.
\end{equation}
Since $K^{0}_{S}$ candidates are reconstructed from $\pi^{+}\pi^{-}$ decays, the detection asymmetry associated with these charged pions cancels.
The remaining reconstruction asymmetry can be written as
\begin{equation}
A_{rec} = A_{CP} + A_{FB} + A_{\epsilon}^{\pi},
\end{equation}
where $A_{FB}$ is an odd function of the cosine of the polar angle of the $D^{+}$ momentum in the CM frame,
and $A_{\epsilon}$ depends on the transverse momentum and polar angle of the $\pi^{+}$ in the laboratory frame.
Assuming the same forward-backward asymmetry for $D^{+}$ and $D^{0}$ mesons, samples of $D^{+}\rightarrow K^{+}\pi^{+}\pi^{+}$ and $D^{0}\rightarrow K^{-}\pi^{+}\pi^{0}$ can be used to correct for $A_{\epsilon}^{\pi}$.
Once $A_{rec}$ has been corrected, the quantities $A_{CP}$ and $\Delta A_{FB}$  can be extracted by adding/subtracting the asymmetry difference in opposite bins of $\cos\theta^{\star}$ using the following relations:
\begin{eqnarray}
A_{CP} &=& \frac{A_{rec}^{cor}(\cos\theta_{D^{+}}) + A_{rec}^{cor}(-\cos\theta_{D^{+}})}{2}, \\
\Delta A_{FB} &=& \frac{A_{rec}^{cor}(\cos\theta_{D^{+}}) - A_{rec}^{cor}(-\cos\theta_{D^{+}})}{2}.
\end{eqnarray}

This measurement is based on 977 fb$^{-1}$ of data collected at or near the $\Upsilon(4S)$ resonance, along with data recorded at the $\Upsilon(1S)$, $\Upsilon(2S)$, $\Upsilon(3S)$, and $\Upsilon(5S)$ resonances, corresponding to 1.74 million reconstructed $D$ decays.
The $D^{\pm}\rightarrow K^{0}_{S}\pi^{\pm}$  signals are parameterized as a sum of Gaussian and bifurcated Gaussian with common mean.
The asymmetry of the negatively and positively charged $D$ yields, as well as their sum, is obtained directly from a simultaneous fit to the distribution of each candidate.
The asymmetry and sum of each yield from the fit are $(-0.146 \pm 0.094)\%$ and $(1738 \pm 2\%)$, respectively, where the errors are statistical only.

The correction to $A_{\epsilon}^{\pi}$ is first extracted using $A_{rec}$ from the $D^{0}\rightarrow K^{-}\pi^{+}\pi^{0}$ sample.
This value is then used to weight each $D^{\pm}\rightarrow K^{\mp}\pi^{\pm}\pi^{\pm}$ candidate with a factor of $(1 \mp A_{rec})$.
The weighted $M(K^{\mp}\pi^{\pm}\pi^{\pm})$ sample is then used to measure $A_{\epsilon}^{\pi}$ in $10 \times 10$ bins of the 2D phase space $(p_{T\pi}, \cos\theta_{\pi})$.

\begin{figure}[htb]
\centering
\includegraphics[height=4in]{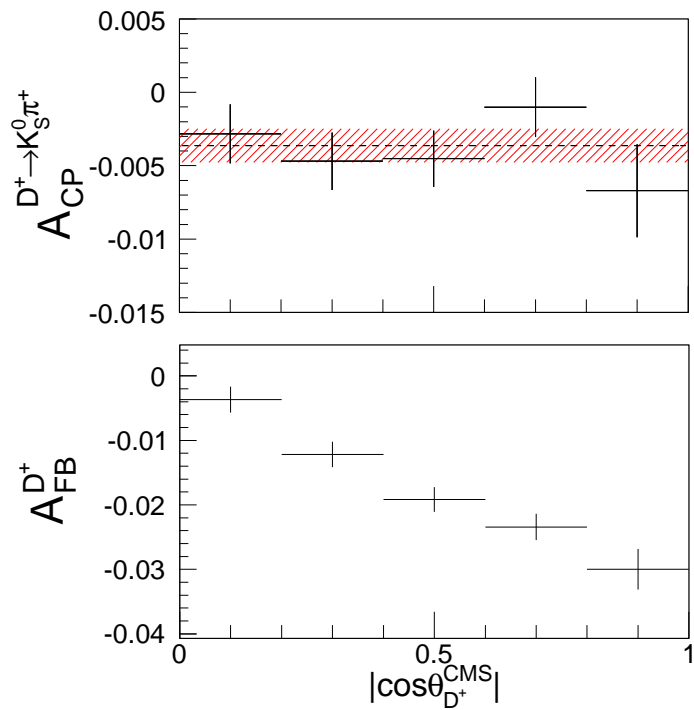}
\caption{Measured $A_{CP}$ (top) and $A_{FB}$ (bottom) values as a function of $\cos\theta_{D^{+}}$. In the top plot, the dashed line is the mean value of $A_{CP}$ while the hatched band is the $\pm1\sigma_{total}$ interval, where $\sigma_{total}$ is the total uncertainty.}
\label{fig:kspi}
\end{figure}

The data samples are then divided into equivalent bins of 2D phase space, whereby each $D^{+}\rightarrow  K^{0}_{S}\pi^{+}$ candidate is weighted by a factor of $(1 \mp A_{\epsilon}^{\pi})$.
The corrected asymmetry is measured using a simultaneous fit of $M(K^{0}_{S}\pi^{+})$ in bins of $\cos\theta_{D^{+}}$.
Fig.~\ref{fig:kspi} shows $A_{CP}$ and $A_{FB}$ as a function of $|\cos\theta_{D^{+}}|$.
The final $CP$ asymmetry measured in the decay is $(-0.363 \pm 0.094 \pm 0.067)\%$, where the first error is statistical and the seccond is systematic.

By subtracting the contribution from neutral kaons, the $CP$ asymmetry from the charm decay is $(-0.018 \pm 0.094 \pm 0.068)\%$~\cite{anal4}.
This value, consistent with zero, represents the most precise measurement of $A_{CP}$ in charm decays to date.
The new world average for this decay mode is listed in Table~\ref{tab:avg}.
\begin{table}[t]
\begin{center}
\begin{tabular}{cc}  
$D^{+}\rightarrow  K^{0}_{S}\pi^{+}$  & $A_{CP}$ \\ \hline
FOCUS  &  $-1.6 \pm 1.5 \pm 0.9$  \\
CLEO  &  $-1.3 \pm 0.7 \pm 0.3$  \\
BaBar  &  $-0.44 \pm 0.13 \pm 0.10$  \\
Belle  &  $-0.363 \pm 0.094 \pm 0.067$  \\
Average~\cite{anal4}  &  $-0.41 \pm 0.09 $  
\end{tabular}
\caption{Summary of measured values and world average for $A_{CP}$ in $D^{+}\rightarrow  K^{0}_{S}\pi^{+}$ decays. The final average is 4.6 standard deviations from zero and consistent with expected $CP$ violation in the neutral kaon system. }
\label{tab:avg}
\end{center}
\end{table}

\section{Conclusion}

In summary, we have reported some of the latest results for the measurement of $CP$ asymmetries in two-body charm decays at Belle. 
Debate continues within the theoretical community as to whether $CP$ violation in the charm sector would indicate new physics, or whether it can be explained within the framework of the Standard Model.  
Although precise theoretical predictions in this sector are very difficult to achieve,
the ongoing search for $CP$ violation in other charm decays remains an exciting one.

\Acknowledgements
The author wishes to thank the organizers and staff of the $5^{\textrm{th}}$ International Workshop on Charm Physics for both a stimulating scientific program and excellent hospitality.


\begin{thebibliography}{99}
\bibitem{Diff}
Corrections for a non-zero $A^{KK}_{\epsilon}$ can be applied iteratively, using $A^{K}_{\epsilon}$ from the previous iteration and Eq.~\ref{eq:akint}. However, negligible changes are found in the asymmetry map after convergence.
\end{thebibliography}
\end{document}